# Using Social Network Analysis to Prevent Money Laundering

Fronzetti Colladon, A. & Remondi, E.

# Using Social Network Analysis to Prevent Money Laundering

Fronzetti Colladon, A. & Remondi, E

**Abstract**

This research explores the opportunities for the application of network analytic techniques to prevent money laundering. We worked on real world data by analyzing the central database of a factoring company, mainly operating in Italy, over a period of 19 months. This database contained the financial operations linked to the factoring business, together with other useful information about the company clients.
We propose a new approach to sort and map relational data and present predictive models – based on network metrics – to assess risk profiles of clients involved in the factoring business. We find that risk profiles can be predicted by using social network metrics. In our dataset, the most dangerous social actors deal with bigger or more frequent financial operations; they are more peripheral in the transactions network; they mediate transactions across different economic sectors and operate in riskier countries or Italian regions. Finally, to spot potential clusters of criminals, we propose a visual analysis of the tacit links existing among different companies who share the same owner or representative. Our findings show the importance of using a network-based approach when looking for suspicious financial operations and potential criminals.

## 1. The problem of Money Laundering

Money laundering, known as the process that transforms the proceeds of crime into clean legitimate assets, is a common phenomenon occurring all over the world. Illegally obtained money is typically "cleaned" thanks to transfers that involve banks or legitimate businesses. This phenomenon is often linked to terrorism, drugs and arms trafficking and exploitation of human beings. It can have strong impact on the economy of a country economy since it helps financing crime and corruption and, as a consequence, makes a country less attractive for foreign private investors: money laundering increases the operational risk of financial transactions and harms the stability of financial institutions (Schott, 2006).

In this study, we analyze the internal transaction database of an Italian factoring company, i.e. a company that buys accounts receivable from third parties at a discounted price, thus repaying the seller with assets that meets its present cash needs and taking the full responsibility of collecting the debts. Past research shows how this kind of business may be linked to asset cloaking (Simser, 2008) or trade based money laundering – with the use, for instance, of fictitious invoicing or price misinterpretation (Naheem, 2015). Therefore, it becomes vital to record and monitor transactions in factoring businesses, and to use the collected data to identify those subjects who have higher risk profiles – those who, with a higher probability, could favor an illicit activity. This is, indeed, the scope of the present research: we tried to develop a model which can be replicated by any factoring company to better profile its clients and the involved third parties, assigning them a risk class. Our models differentiate from past studies since we use a relational approach (Wasserman & Faust, 1994) and we present new ways to sort out relationship data to build several interaction networks. The results, which could be integrated with other approaches, were tested on real world data.

A large body of literature addressed the problem of money laundering (Reuter & Truman, 2004), many times ignoring the fact that illegal transactions require the interaction between social actors; by contrast, we suggest to focus not just on the individual attributes of these actors, but to include the study of their relationship networks, by means of Social Network Analysis (Dreżewski, Sepielak, & Filipkowski, 2015; Sparrow, 1991) – a well-known set of methods and tools for the analysis of social dynamics (Wasserman & Faust, 1994).

In Section 1.1, we started by presenting some guidelines extracted from the most recent European directives, which we took as a preliminary reference to identify the most important variables to include in our models. In Section 2, we describe previous attempts and methods to detect and stop the circulation of dirty money. In Section 3, we present our case study and research model; in Sections 4 and 5, we present and discuss our findings.

### 1.1. European/Italian Guidelines and Recording of Financial Data

Since our case study takes into consideration an Italian factoring company, we briefly recall the content of the main European ad Italian guidelines that we took as a reference for setting up an efficient data analysis methodology that could help in identifying those people or companies with a higher risk profile, in terms of illicit transactions. We started with the examination of the directive 2005/60/EU of the European Parliament (2005) implemented by the Italian parliament in 2007. Thanks to this implementation, financial intermediaries operating in Italy now have the duty to keep track of every financial operation they carry out and to record them in a specific database, accessible in case of need by the Financial Intelligence Unit (FIU) of the Bank of Italy. This represents a big step forward, if compared with how transaction data was before stored by some of the smaller companies: the new database has to meet minimum technical requirements in terms of quality and type of information stored. In addition, credit and financial institutions can no longer keep anonymous accounts and need to apply customer due diligence measures – also reporting suspicious activities, like account transaction that exceed a certain amount – with exceptions and reinforcements depending on the specific cases.

For a factoring company, customer due diligence implies gathering and recording the identity of customers and, when these are firms, the recording of the owner data and of the person responsible for the process of money transfer, or the financial operation. The nature, the amount and the scope of the transactions are also investigated and controlled over time. In other words, the central database of each company should contain all the relevant information about transactions, accounts and business relationships. Aggregated data about operations is transmitted monthly to the FIU which carries out a more systemic analysis at a national level. In addition, the FIU must be informed every time an institution suspects money laundering or terrorist financing. Accordingly, one of the main issues remains to find an efficient way to identify suspicious actors and transactions. This identification should be the outcome of a process that cannot rely only on individual judgements, but should – at least for a part – be automated. National guidelines suggest taking into account a list of criteria referred to financial actors, such as their legal status, core activity, possibly suspicious behavior and geographic area. As regards each operation, one should focus the attention on the typology, the amount, the reason, the frequency and the consistency with previous activity and geographic area. What we propose is to enrich this set of information by studying social actors and their transaction data (i.e. their links), not as single entities, but from a network perspective, to benefit from the power of relational data.

## 2. Detecting Money Laundering

In the past decades, most of the suspicious activities were identified by considering anomalies in the regular transactions flows of a client: behavior trends were compared over time to spot unexpected turns. A large body of literature addressed the topic of contrasting money laundering, with several studies proposing additional guidelines and methodologies (e.g., Reuter & Truman, 2004; Schott, 2006; e.g., Sharman, 2008).

When data is big and a deeper knowledge of the clients is sometimes missing – consider for instance regular bank agencies, which may have thousands of customers, mostly operating online or interacting sporadically with several different cashiers – then a computer-based approach seems a reasonable solution to extract relevant information. Data mining is now largely applied in business and for managing information systems (Bose & Mahapatra, 2001; Turban, Sharda, & Delen, 2015), not only with the objective of detecting fraud events, but also with the intent of preventing them.

Rule-based classification approaches are frequent. Rajput and colleagues (2014), for instance, presented an ontology-based expert system to detect suspicious transactions, using the Semantic Web Rule Language. Khanuja and Adane (2014) combined the analysis of forensic databases with the Dempster Shefer Theory of Evidence to automatically generate reports of suspicious actors. Other studies used Bayesian approaches either to assign a risk score to customer behavior, also taking into account the evolution of such behavior (Khan, Larik, Rajput, & Haider, 2013), or, more generally, to combine Bayesian rules with database information about past operation history (Panigrahi, Kundu, Sural, & Majumdar, 2009), outperforming other fuzzy techniques.

Machine learning algorithms were also applied with the same general purpose. Tang and Yin (2005) developed a classification system based on support vector machine in order to handle large amounts of data and reach more accurate predictions than those based on traditional rules.

Castellón González and Velásquez (2013) combined decision trees, neural networks, Bayesian networks and clustering algorithms to detect false invoicing of taxpayers. Clustering algorithms were also explored by Wang and Dong (2009) – who introduced a new dissimilarity metric – and by Roith and Patel (2015) who reviewed various data mining techniques and Anti-Money Laundering (AML) detection methods, arguing that clustering techniques are the best solution. By contrast, Axelsson and Lopez-Rojas (2012) discussed some of the limitations of the machine learning approach – such as the problem of the identification of parameters that are data dependent, with sometimes limited adaptability and scalability, the risk of misclassification, the class unbalance problem, or the difficulties in the interpretation of the underlying behavioral patterns – and proposed Multi-Agent Based Simulation as a valuable integration, or alternative. Other agent-based solutions to support decision making in AML were discussed by Gao and Xu (2009).

In addition, we found studies which tried to identify illegal fund transferring through the combination of transaction flow information and the analysis of customer behavior (Kharote & Kshirsagar, 2014). Similarly, Perez and Lavalle (2011) based their methodology on the detection of outliers in transactions, modelling customer behavior in two stages: first, they built user models on historical data and then they compared new transactions against these models. This way the authors stressed the advantage of studying individual patterns, without the need of defining a general rule to be applied to all actors or financial operations. However, we see a main limitation laying in the model requirement of historical data, making more difficult to identify illicit operations made by newcomers.

It is now evident that a large body of literature is addressing the topic of anti-money laundering and fraud detection. Some scholars produced an organized review of the major methods, trying to reduce the complexity in the interpretation of single investigations (Z. Gao & Ye, 2007). Ngai et al. (2011) produced one of the first systemic literature reviews, structuring the different contributes depending on the fraudulent activity they were addressing and on the data mining application classes (classification, clustering, prediction, outlier detection, regression, visualization) and techniques (logistic models, neural networks, decision trees, Bayesian belief networks). Anomaly detection techniques in financial domains were recently surveyed by Ahmed at al. (2015); they specifically focused on clustering based techniques, discussing their assumptions and making a comparison from different perspectives; they also stressed the importance of working on real world data to validate model results. Accessing real data from financial organizations is often very difficult, due to privacy reasons and competition. We succeeded in collecting such kind of data and in validating our models accordingly.

A last set of studies has focused the attention on relational data, in order to reveal new patterns and surpass some of the limits of the traditional approaches. There is a tradeoff between the need for financial institutions to rely on rules that can be easily applied and understood and the request for more sophisticated methodologies, which should be difficult to dodge, not relying on pre-established and well-known rules that might be easily eluded. Consistently, we propose to include social network metrics (Wasserman & Faust, 1994) in the monitoring system, with the advantage of extending the set of measures that might complete a client's risk profile and to offer a set of parameters very difficult to hack. Even if money launders discovered their network is monitored, it would be much more difficult for them to change their behavior to the point of influencing more complex indicators, such as betweenness centrality (Freeman, 1979) and its oscillations over time. A Social Network Analysis (SNA) approach has already been used to detect and disrupt covert networks of criminals and terrorists (Carley, 2006; Everton, 2012; Roberts & Everton, 2011). Authors like Sparrow (1991) proved the advantage of representing data in link form, to answer questions such as "who is central in the criminal network?", "Which names look like an alias?", "What are the anomalies in cash flow patterns?", or "Is there an evidence of smurfing?". More recently, Drezewski, Sepielak and Filipkowski (2015) extracted relational data from bank statements and court registers to specifically support money laundering detection, through the assessment of the role of individual nodes, considering centrality measures and network clustering. Other contributes came from Gao and Ye (2007) who proposed a data mining framework; they reorganized findings from previous studies also stressing the importance of link analysis. Didimo, Liotta, Montecchiani and Palladino (2011) developed a system for the visual analysis of financial networks, which relies on automatic *k-core* clustering and measures a range of centrality scores for each node in the graph. Network visual analysis can be very useful for AML, but it loses its power when the graph is extremely big. Moreover, we maintain the importance of mapping different kinds of relationships, resulting in several possible networks. While analyzing these networks, many centrality measures, and hundreds of clustering algorithms, are possible. In presenting our results, we show the metrics which better performed in our case study and which can overcome the limits of the visualization approach for large graphs.

Despite the large body of literature addressing the topic of AML, it is still unclear which set of methods might better perform; many scholars keep defending their findings still being unable to prove the preeminence of their choice. It seems that neither adopting a single strategy nor just considering the European guidelines is sufficient to fully prevent illicit transactions. Accordingly, we believe that a mixed approach, combining different methodologies that include relational data, could outperform individual findings. However, proving this is not in the scope of our investigation, but rather a proposal for future research. Here we intend to focus the attention on the value of relational data and to propose new applications of metrics that could be integrated in a more comprehensive AML platform. Even if the use of Social Network Analysis in AML is relatively new and not yet fully explored, we find network metrics extremely useful when performing fraud-risk assessments. Some previous research following the SNA approach proved to be extremely data dependent, often requiring the integration of multiple sources; accessing multiple data is desirable, but not always possible. Our method proves its highest utility when almost on-time decisions are required and a single data source is available, i.e. the database of a medium-large financial institution (it could be a bank or a factoring company).

## 3. Case Study and Research Design

In order to test the effectiveness of a network analysis approach in preventing money laundering, we analyzed the central database of a medium-large factoring company in Italy, where the financial operations from November 2013 to June 2015 are recorded. According to the Italian law, each financial institution has to record all the transactions amounting to at least 15,000 euros and smaller transactions in case their aggregated value is bigger or equal to 15,000 euros (to face the problem of smurfing). This company is well established and has actively operated for more than 20 years. Due to privacy reasons we cannot reveal the company name and we had to anonymize all the collected data.

We recall the factoring business as the process where the factoring company (*factor*) purchases a receivable (usually an invoice) at a discounted price from a *seller,* who in this way will immediately satisfy its cash needs. Later on, the factor will receive a payment from the *debtor* who is the one who has a financial liability that requires him or her to make a payment to the owner of the invoice. The process is presented in Figure 1.

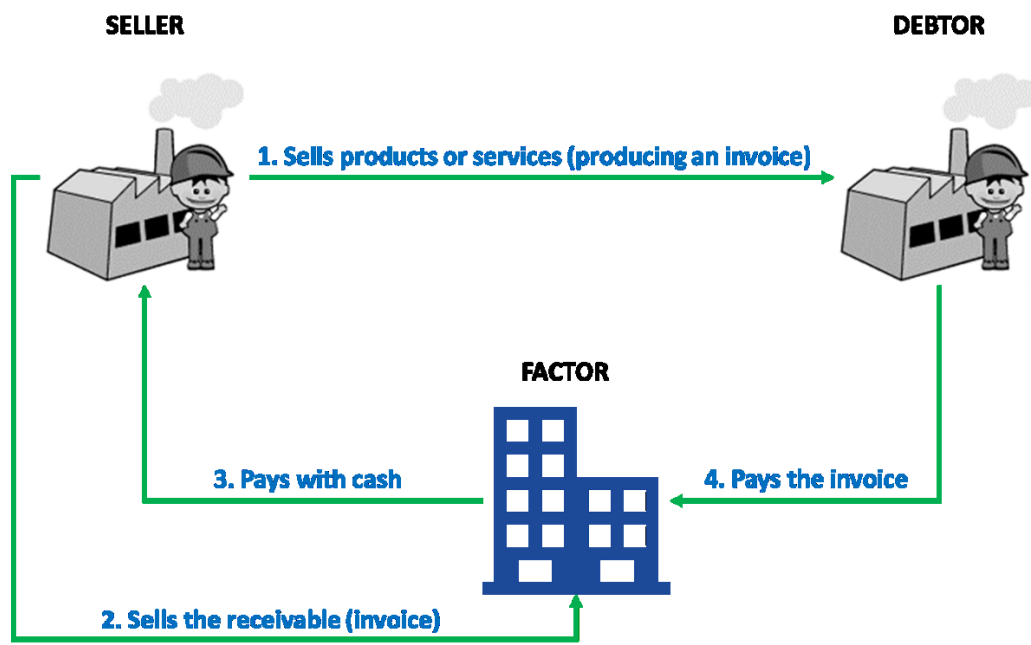

**Figure1.** The factoring process.

### 3.1. Networks Extraction and Risk Factors

To profile network nodes and identify those with a higher risk of fraud, we analyzed all the economic transactions concerning the factoring business. Such analysis is also useful to visually give evidence to the relationships existing between sellers and debtors, drawing the attention on suspicious cases where, for instance, the seller and the debtor share the same beneficial owner. A part of illicit financial operations, indeed, is linked to cases where goods or services are not really supplied from the seller to the debtor, leading to fictitious invoicing or price misinterpretation (i.e. inflated prices). This as a quite common way to clear money coming from illegal sources or to evade taxes; Castellón González and Velásquez (2013, p. 1427), for instance, reported that "evasion by false invoicing has historically represented between 15% and 25% of total VAT evasion, increasing significantly in years of economic crisis"; in the data analyzed by Irwin, Choo and Liu (2012) fictitious invoicing was used in the 10% of cases of money laundering and fictitious sales/purchases in the 9% of cases.

With respect to points 3 and 4 of Figure 1, we focused our analysis on the money transfers made from the debtor to the factor (point 4): these are the operations used to put dirty money in the clean circuit. Particular attention was then reserved to those money wirings involving companies operating abroad. In addition to the amount of each payment, the database contains information about the seller and the debtors associated to that payment. Consistently, we started building a network of transactions where we directly linked sellers and debtors, to track their connections, without representing the factoring company (Figure 2). It includes 559 nodes (sellers/debtors and nodes with a double role) and 33,670 links (representing money transfers in Italy and abroad). It is worth noting that some nodes might share a multiple role, being sellers in some cases and debtors in other cases. This is well represented by nodes having both outgoing and incoming ties.

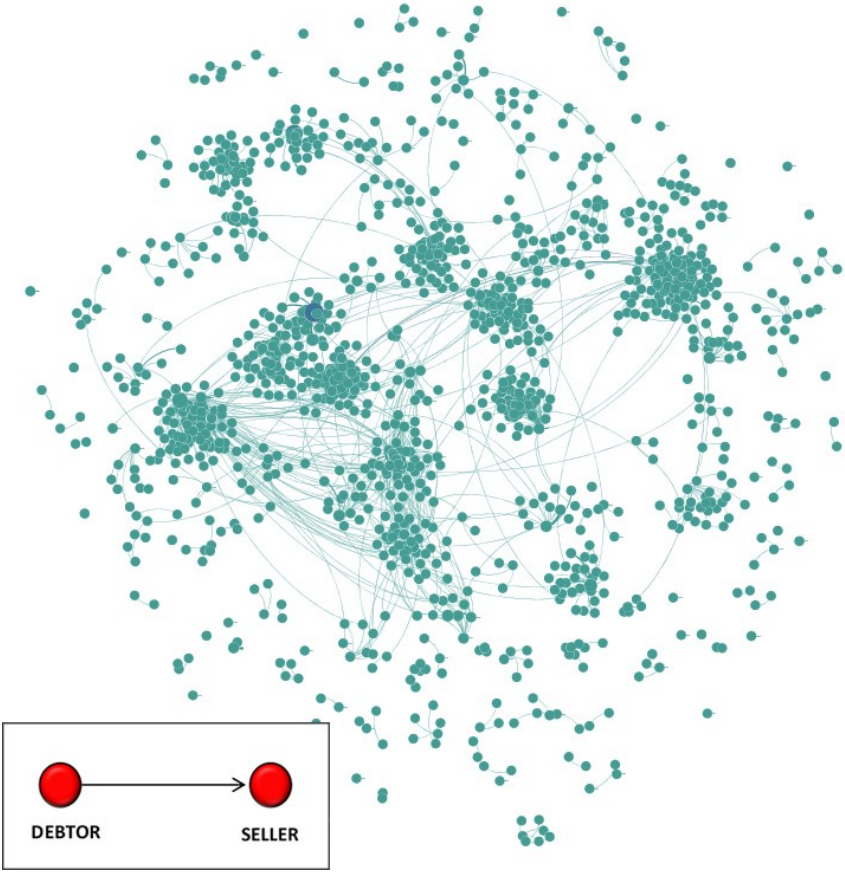

**Figure 2.** The transactions network

Starting from this complete representation, several other networks are possible; to build them we filtered the original graph considering some of the major risk factors reported by the Financial Intelligence Unit of the Bank of Italy (2014). We specifically paid attention to the geographical area and the economic sector of the seller and the debtor, and to the amount of financial operations. For each of these three risk factors, we constructed a separate network. Such a choice allows us to extract more comprehensive information from the database than if we just analyzed the full network once. To put it in other words, the analysis of a separate network for each risk factor allowed us to evaluate the individual contribute of each of them and to calculate more network metrics. Moreover, this way we can avoid mixing conceptually different elements.

*Transactions Network*. Illicit transactions can be part of a large money transfer or can be hidden within a greater amount of completely legal transactions. In both cases, one could expect that criminals will try to hide illicit money transfers choosing firms that usually deal with larger and/or more frequent transactions. The database of the factoring company does not contain sufficient information to distinguish between legal and illegal transactions, nor is this information known at the time of the analysis. Therefore, our aim is not to identify specific illegal money transfers as they occur. Nonetheless, we maintain the importance of considering the frequency and the amount of money transfers between each pair of actors – even when one is not aware of their legality – to identify suspicious clients. Consistently, we built a Transaction Network where each arc is weighted with a score ranging from 1 to 3, depending on the associated transaction amount. Specifically, we assigned a score of 1 to all the transactions with an amount below 50,000 euros, a score of 2 if the amount was between 50,000 and 249,999 euros and a score of 3 for higher amounts. Operating this binning leads, in our case study, to a much better performance of the predictive models presented in Section 4. Other thresholds are also possible – as well as weighting the arcs with the nominal amount of each transactions – depending on the specific case, so we leave this final choice to the analyst. This method, combined with the network metrics presented in Section 3.2, also allows to partially address the problem of smurfing – the execution of multiple smaller financial transactions to avoid the creation of records required by law – since one can also consider the aggregated value of multiple operations existing between two nodes, again with the limitation of not having sufficient information to distinguish between legal and illegal money transfers. However, the scope of our research is to evaluate the risk profiles associated to each client of the factoring company and not to single transactions.

In the next two graphs, the risk level does not depend directly on a link property (like its amount), but it is determined by the attributes of the nodes connected by that link. Accordingly, we calculate a risk level for each node, depending on the geographical area or on its economic sector; subsequently, we assign a risk score to each link, calculated as the average of the risk level of the nodes it connects. Other scoring techniques are possible – like taking the maximum instead of the average – but we found the average is the most effective choice in our case study.

*Economic Sector Network*. For the coding of economic sectors we referred to the ATECO 2007[1] national classification of economic activities, developed by the Italian National Institute of Statistics (ISTAT), in accordance with the European nomenclature (REGULATION (EC) No 1893/2006, 2006). Seller and debtor companies – all associated with an ATECO code in the factor database – where divided into two classes, depending on their economic sector being classified as high or low risk of fraud. This classification was made by two experts working together for the factoring company: they analyzed the national reports produced by the FIU and also considered the number of suspicious and illicit transactions. Specifically, we here list all the sectors classified as high risk (all the remaining sectors were classified as low risk): public procurement, construction, ferrous materials (production and sales), car sales, road haulage, high tech goods (sales), cleaning and maintenance services, consultancy services, precious metals and gemstones (production and sales), artworks (sales), oil and grain wholesale, advertising, online IT services, supermarket, betting and gambling, weapons and explosive materials, waste management, waste water treatment. Several classification techniques to identify high risk sectors are possible – machine learning or clustering algorithms might come useful to the purpose. However, it is not in the scope of our research to find or defend the best method, nor is our knowledge of business sectors large enough to propose a classification ourselves: our contribute here was more about testing the effectiveness of the classification made by the company experts. We also suggest revising each classification year by year, according to the updates of national reports.

---

[1] Details available at: http://en.istat.it/strumenti/definizioni/ateco/ (Retrieved Jun 23rd, 2016).

*Geographical Area Network.* We generated a classification of geographical risk associated to Italian regions and foreign states. At a national level, we considered, for each region, the percentage of crimes per province as reported by the ISTAT; the number of suspicious operations, taken from the FIU reports; the fact of being classified as a region with a bigger presence of Mafia. For each of these three measures, we calculated a partial risk score (ordinal, ranging from 1 to 3) according to whether the Italian region was approximately in the lowest 30%, middle 40%, or highest 30% of the percentile estimates of each measure. By averaging these three partial risk factors, we obtained a geographical risk score for each region. Subsequently, we weighted each arc (transaction) in the network with a risk score, calculated as the average of the risk scores associated to the regions where the seller and the debtor operated. With regards to those sellers/debtors operating in foreign states, we proceeded in the same way, but considering different measures. Specifically, we accessed several information sources to determine risk levels: the "white list" diffused by the Italian Ministry of Economy and Finance, which includes the countries complying with AML procedures that are consistent with the EU guidelines; the list of tax havens; the country level of compliance to OCSE exchange of tax agreements; the Corruption Perception Index[2]; the list of jurisdictions that have strategic deficiencies and pose a risk to the international financial system (FATF Public Statement). However, transactions involving foreign states are almost negligible in our dataset, representing less than 1% of the total money transfers.

We created a last network – which we called the *Tacit Link Network* – to take into account a risk factor not sufficiently stressed in the common guidelines: the implicit connection among clients who have a factoring relationship and share the same beneficial owner or person delegated to carry out a financial operation. We claim such analysis is vital to spot tacit, potentially illicit, agreements. By *owner* we intend the single person who is making the financial operation on his/her name, or alternatively the person managing the company who is making a specific operation; by *representatives* we mean those people (usually employees) who are authorized to operate on behalf of a company. In this new undirected graph, we linked two nodes if they shared the same owner or representative. We found 90 connected nodes and, after having removed the isolates, we carried out a visual analysis as described in Section 4.

---

[2] Details available on http://www.transparency.org/research/cpi/overview (Retrieved Jun 23rd, 2016)

Starting from the first three graphs, we calculated the metrics presented in the next section. Before operating the calculus we removed the loops (cases where nodes link to themselves, due to missing data about the debtor). Moreover, we transformed the graphs referred to the economic sectors and the geographical areas, by removing multiple arcs and taking the average of their weights. This last task is not mandatory: one could keep multiple arcs for all networks, or sum their values, or take their maximum value. We leave this decision to the analyst, given that, in our case, a simplification using the average provided the best final results. Lastly, we focused on high risk operations, by filtering out medium and low risk links; this, again, to improve model performance.

### 3.2. Network and Control Variables

Using a Social Network Analysis approach, several strategies are possible to identify nodes with a higher risk of fraud. In order to make our method easier to understand and replicate, we referred to some of the most common measures of centrality – *in-degree, out-degree, all-degree, closeness* and *betweenness centrality* – as a way to identify the most relevant social actors (Freeman, 1979; Wasserman & Faust, 1994). These measures are commonly used by network analysts and they are well described in the work of Freeman (1979), Wasserman and Faust (1994), or in other Social Network Analysis manuals (e.g., De Nooy, Mrvar, & Batagelj, 2011). We start from the notation used in these sources to briefly recall the main formulas and calculus procedures. In general, we can consider an oriented graph made of a set of *n* nodes (social actors) – referred as G = {$g_1$, $g_2$, $g_3$ … $g_n$} – and of a set of *m* oriented arcs connecting these nodes – referred as A = {$a_1$, $a_2$, $a_3$ … $a_m$}. This graph can be represented by a sociomatrix X made of *n* rows and columns, where the element $x_{ij}$ positioned at the row *i* and column *j* is bigger than 0 if, and only if, there is an arc ($a_{ij}$) originating from the node $g_i$ and terminating at the node $g_j$. When the elements of X are bigger than zero, they represent the weight of the arcs in the graph ($x_{ij}$). If arcs are not weighted, we consider a dichotomized sociomatrix where each element can only assume the value 0 or 1.

*(Weighted) In-degree*. This measure counts the number of incoming ties, i.e. the number of times a client acts as a seller. In the weighted version, incoming arcs values are summed up, to calculate the risk associated to the total number of operations in the database. The (weighted) in-degree for the generic node $g_i$ is represented by the equation:

$$D_I(g_i) = \sum_{j=1}^{n} x_{ji}$$

*(Weighted) Out-degree.* This measure counts the number of outgoing ties, i.e. the number of times a client acts as a debtor and can be weighted as the in-degree variable. The (weighted) out-degree for the generic node $g_i$ is represented by the equation:

$$D_O(g_i) = \sum_{j=1}^{n} x_{ij}$$

*(Weighted) All-degree.* With this measure, we count the number of ties connecting a node to the others, regardless of their directionality. In the weighted version, a risk factor is calculated as the sum of the weights of all the arcs connected to one node. This measure substitutes in-degree and out-degree on undirected networks. The (weighted) all-degree for the generic node $g_i$ is represented by the equation:

$$D_A(g_i) = \sum_{j=1}^{n} (x_{ij} + x_{ji})$$

*Closeness.* This variable expresses the inverse of distance of a node from all others in the network, considering the shortest paths that connects each couple of nodes. It is often used as a proxy of the speed by which a social actor can reach the others. If we let $d(g_i, g_j)$ be the number of arcs in the shortest path linking the nodes $g_i$ and $g_j$ we can express the closeness centrality of the node $g_i$ with the following equation:

$$C_C(g_i) = \frac{1}{\sum_{j=1}^{n} d(g_i + g_j)}$$

where $\sum_{j=1}^{n} d(g_i + g_j)$ is the distance of the node $g_i$ from all the other nodes in the graph (taking this sum over all $j \neq i$ ). This index can be standardized multiplying its equation by $g - 1$, to allow the comparison across networks of different sizes (Wasserman & Faust, 1994).

*Betweenness*. This measure is higher when a node is more frequently in-between the shortest paths that connect every other couple of nodes. To calculate the betweenness centrality of the node $g_i$ we use the following equation:

$$C_B(g_i) = \frac{\sum_{j<k} d_{jk}(g_i)}{d_{jk}}$$

where $d_{jk}$ is the number of shortest paths linking the generic couple of nodes $g_j$ and $g_k$, and $d_{jk}(g_i)$ is the number of that paths which contain the node $g_i$. This measure can be standardized dividing it by $[(n-1)(n-2)/2]$, which is the total number of pairs of actors not including $g_i$ (Wasserman & Faust, 1994).

Another important measure, the network constraint, has been introduced by Burt (1995) when describing the concept of *structural holes*. To give an example, we consider three nodes A, B and C, where A is linked to B and C, but a link between these last two is missing. That missing link is called "structural hole" and gives a social advantage to A who can adopt a "dividi et impera" strategy, like mediating a trade between B and C who cannot communicate directly, thus benefiting from the application of an additional price charge. In this sense, we say that A is less "constrained" by its ego-network and can benefit from the structural hole existing between B and C. It has been proved that brokerage across structural holes can have several advantages, such as more possibilities of being promoted, listened and positively evaluated in a business context (Burt, 2004).

*Network Constraint*. It measures the value of the network constraint, for each node, as presented by Burt (1995) and furtherly described in the book illustrating the software Pajek (De Nooy et al., 2011) which we used to calculate all our network variables. The less the structural holes in an ego-network, the more the node is constrained to the group rules.

Lastly, we considered a control variable, linked to the presence of unrecorded data, that could help predicting the risk profile of clients. There is a limited number of operations where the beneficial owner, the representative, or the debtor are not recorded in the database and so missing. Even if missing data might come from technical reasons, from error in communication between different operating systems, or from errors in the data input, we believe it is important to specifically control for such cases, to see if there is a recognizable pattern behind them.

*Missing Id.* It counts the number of times a seller is associated to financial operations where the owner, the representative, or the debtor are missing.

Controlling for age and size of seller and debtor companies could be an important addition to the present study and a proposal for future research. Nonetheless, we could not collect such information, as it was not recorded in our database.

From the analysis of high risk transactions and network metrics, we try to predict those clients that could be more likely involved in illicit operations. To evaluate risk profiles of the companies registered in the database, we examined Italian court records to check whether either their owners or representatives were involved in anti-money laundering investigations and trials at the time of data collection or in the previous three years. Accordingly, we created a binary dependent variable (named *High Risk*) that has the value of one if either the owner or the representatives were involved in legal trials and has the value of zero otherwise. Collecting court data was not always possible and entailed the recombination of fragmented data sources, thus having the possibility to gather complete data for 288 cases over the total sample. The reduction in the number of complete observations was also partially due to missing data about the owner or the representative. We still considered the full sample when calculating network metrics.

We hypothesize that more central nodes – with a higher number of connections, a more frequent involvement in economic transactions and a larger presence in the paths that connect other nodes – have a higher risk profile, suggesting a deeper investigation. We claim a higher risk should also be associated to nodes with a lower network constraint. In this study, we explore the significance of such contributes. We seek for the most informative metrics applying them to all the presented networks, filtered and recombined as described in Section 3.1. Centrality measures and the analysis of structural holes have already been used in the past to detect criminals (Ferrara, De Meo, Catanese, & Fiumara, 2014; Sparrow, 1991) or, more in general, to disrupt dark networks (Carley, 2006; Everton, 2012; Scott & Carrington, 2011).

## 4. Results

Our findings confirm that social network metrics are important elements to consider when assessing risk profiles. We also prove that working with multiple networks improves the informative power of the single metrics. Correlation coefficients of our predictors with the dependent variable are presented in Table 1.

|   |   | 1 | 2 | 3 | 4 | 5 | 6 | 7 | 8 | 9 | 10 | 11 | 12 | 13 | 14 | 15 | 16 | 17 | 18 | 19 | 20 |
|---|---|---|---|---|---|---|---|---|---|---|---|---|---|---|---|---|---|---|---|---|---|
| 1 | High Risk | 1 | | | | | | | | | | | | | | | | | | | |
| 2 | Missing Id | .166** | 1 | | | | | | | | | | | | | | | | | | |
| 3 | In-degree Geographical Area | .249** | .100 | 1 | | | | | | | | | | | | | | | | | |
| 4 | Out-degree Geographical Area | .312** | .156** | .282** | 1 | | | | | | | | | | | | | | | | |
| 5 | All-degree Geographical Area | .257** | .099 | .980** | .378** | 1 | | | | | | | | | | | | | | | |
| 6 | Closeness Geographical Area | .049 | .089 | .456** | -.260** | .404** | 1 | | | | | | | | | | | | | | |
| 7 | Betwenness Geographical Area | .296** | .512** | .150* | .262** | .167** | .099 | 1 | | | | | | | | | | | | | |
| 8 | Network Constraint Geographical Area | -.108 | -.094 | -.184** | -.669** | -.248** | .412** | -.153** | 1 | | | | | | | | | | | | |
| 9 | Weighted In-degree Transactions | .240** | .103 | .090 | -.222** | .054 | .352** | .159** | .263** | 1 | | | | | | | | | | | |
| 10 | Weighted Out-degree Transactions | .203** | .288** | .153** | .425** | .214** | .026 | .546** | -.052 | -.068 | 1 | | | | | | | | | | |
| 11 | Weighted All-degree Transactions | .314** | .106 | .109 | -.064 | .089 | .313** | .195** | .224** | .910** | .169** | 1 | | | | | | | | | |
| 12 | Closeness Transactions | -.053 | .091 | .095 | -.338** | .040 | .761** | .101 | .442** | .436** | -.026 | .387** | 1 | | | | | | | | |
| 13 | Betwenness Transactions | .254** | .459** | .173** | .222** | .187** | .094 | .893** | -.172** | .187** | .611** | .218** | .113 | 1 | | | | | | | |
| 14 | Network Constraint Transactions | -.220** | -.156** | -.301** | -.235** | -.301** | -.137* | -.279** | .186** | -.500** | -.233** | -.569** | -.206** | -.313** | 1 | | | | | | |
| 15 | In-degree Economic Sector | .236** | -.013 | .213** | .394** | .250** | -.044 | .043 | -.230** | .006 | .016 | -.001 | -.081 | .028 | -.183** | 1 | | | | | |
| 16 | Out-degree Economic Sector | .326** | -.001 | .205** | .608** | .266** | -.077 | .184** | -.359** | .002 | .121* | .043 | -.109 | .157** | -.195** | .729** | 1 | | | | |
| 17 | All-degree Economic Sector | .259** | -.012 | .225** | .412** | .261** | -.042 | .129* | -.241** | .016 | .063 | .016 | -.079 | .105 | -.205** | .985** | .756** | 1 | | | |
| 18 | Closeness Economic Sector | .088 | -.017 | .169** | .043 | .161** | .047 | .201** | -.003 | .115 | .072 | .087 | .031 | .173** | -.226** | .671** | .171** | .701** | 1 | | |
| 19 | Betwenness Economic Sector | .254** | .459** | .173** | .222** | .187** | .094 | .893** | -.172** | .187** | .611** | .218** | .113 | 1.000** | -.313** | .028 | .157** | .105 | .173** | 1 | |
| 20 | Network Constraint Economic Sector | -.270** | -.001 | -.137* | -.531** | -.197** | .108 | .031 | .333** | .099 | -.013 | .069 | .143* | .034 | .050 | -.672** | -.876** | -.662** | .071 | .034 | 1 |

**p<.01; *p<.05.

**Table 1.** Pearson correlation coefficients.

Degree centrality emerges as an important predictor, having more central nodes associated with risky profiles, in all networks. The same applies to betweenness centrality which is always positively correlated with high risk. Closeness centrality, on the other hand, is not significant, even if this predictor shows some valence in the logit models (Table 2). Network constraint in the Economic Sector Network turns out to be another important variable, suggesting that more open ego-networks should capture the attention of the analyst. Controlling for missing data also seems to be fairly important. The high correlations between the in- and out-degree variables and their corresponding all-degree are normal, being the all-degree the sum of in- and out-degree scores. It is also worth noting that, in our specific case, the betweenness centrality score in the Transactions Network has the same ranking of betweenness in the Economic Sector Network. This can happen since all the proposed graphs originate from the same core (the transactions network).

To extend our first results, we tested several logit models to assess the predictive power of the SNA metrics, for all the networks described in Section 3.1. Main outcomes are reported in Table 2 (for the sake of readability, we avoid including all the non-significant predictors and we limit to the best final models, also avoiding cases of collinearity).

| Variables | Model 1 | Model 2 | Model 3 | Model 4 |
|---|---|---|---|---|
| Missing Id | .080*** | | | .041* |
| In-degree Geographical Area | | .551*** | | .318* |
| Network Constraint Economic Sector | | -1.815** | | -2.254** |
| Weighted All-degree Transactions | | | .790*** | .814*** |
| Closeness Transactions | | | -.465** | -.543** |
| Constant | -2.157*** | -2.025** | -2.955*** | -2.063** |
| McFadden's $R^2$ | .126 | .132 | .167 | .327 |
| N | 288 | 288 | 288 | 288 |
| AIC | 206.802 | 207.402 | 199.436 | 168.171 |
| BIC | 214.128 | 218.391 | 210.425 | 190.150 |

***p<.001;**p<.01; *p<.05

**Table 2.** Identifying high-risk profiles

The Table shows that clients associated to transactions with missing information – about the owner, the representative, or the counterpart – are more likely to have a high-risk profile. In addition, three networks play a major role when assessing risk profiles: those constructed considering the economic sector, the geographical areas of activity, and the total amount of financial operations. As far as the network metrics are concerned, degree centrality, closeness centrality and the network constraint proved their significance. By contrast, we surprisingly notice a marginal role of betweenness centrality, which is significant in the correlations, but loses its importance in the predictive models. Model 2 shows that companies that more frequently buy products or services from firms operating in high risk geographical areas are more probably involved in illicit operations. Moreover, companies with a lower network constraint, spanning their activities between riskier economic sectors, should be considered more suspicious and deserve more attention. In Model 3, we appreciate the importance of studying the transaction network: the higher the total amount of transactions, the higher the risk of fraud. The same risk decreases when closeness is higher, indicating that less isolated clients, more central in the network, are less likely to act irregularly. This last finding partially contrasts our hypothesis that more central nodes have higher risk profiles; nonetheless, maintaining a more central position in a network where criminals are a minority might also push companies to ask honestly and keep a good reputation. However, the effect of closeness is rather small if compared with the other variables. In the final model, we put predictors together – obtaining a Mc Fadden's R-Squared of 0.327 and significant reductions in AIC and BIC scores, which demonstrate a good predictive power and prove the validity of our approach. Consistently with our models, we maintain that the financial operations of those clients who can be associated to higher risk profiles should be signaled to the local authorities, to allow a more in-depth investigation at a national level.

As a last step, we analyzed the non-isolate 90 nodes in the Tacit Link Network. Not surprisingly, this network is fairly sparse, with fewer link information. However, from a visual analysis we immediately notice that nodes are well divided into separate cluster. To investigate the groups structure, and the presence of network clusters, is common in this field (e.g., Carley, 2006; Everton, 2012; Sparrow, 1991). In our case study, the different groups are so evident that we did not need to apply clustering algorithms. This network is presented in Figure 3, having excluded the large proportion of isolates.

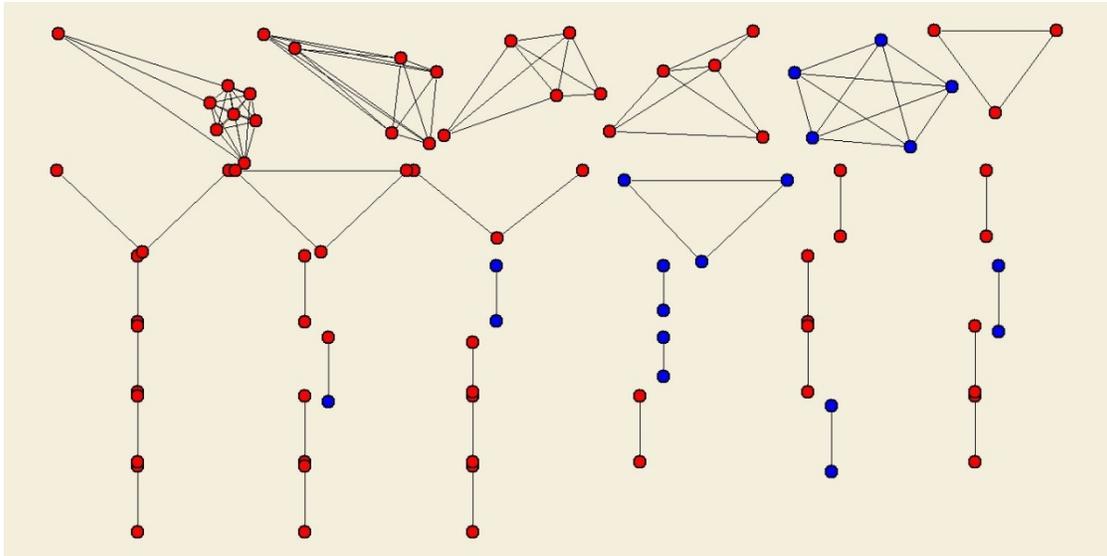

**Figure3.** Tacit Link Network (no isolates)

Blue nodes are those clients who were involved in AML or crime financing trials. As we see from the picture, they are almost perfectly clustered. Accordingly, the fact that a non-isolate node becomes blue, at a certain moment in time, should act as a wake-up call about the other nodes in its cluster, which could get soon involved in legal trials, presenting a higher risk profile. This last part of the analysis is mostly exploratory and was conceived as an easy and immediate way to point the analyst attention to possible cases of suspicious clients. Obviously, the fact that a company in a cluster gets involved in criminal proceedings does not always imply that all the other companies in that cluster will be involved in the same proceedings – some companies could, for instance, operate in different sectors and carry out totally independent activities, even when sharing the same owner.

## 5. Discussion and Conclusions

The key contribution of this paper is to emphasize the role of social network metrics and introduce new network mapping techniques, not traditionally associated with anti-money laundering practices. Financial companies with the duty of identifying and reporting suspicious operations and behaviors could benefit from our models, and possibly integrate them with other techniques. Consistently with previous studies (e.g., Didimo et al., 2011; Z. Gao & Ye, 2007;

Sparrow, 1991), we defend the more informative power of relational data, when compared with the study of the single attributes of social units.

In our case study, we analyzed the central database of a medium-large factoring company operating in Italy; we started by mapping 4 different kinds of relational graphs – the first, to take into account the operational risks associated to the economic sectors of activities; the second, to consider the risk associated to geographical areas; the third, to study the amount of transactions and the fourth, to bring to light tacit, potentially harmful, links between different companies sharing the same owner or representatives. We filtered these graphs to focus on transactions at higher risk and we kept risk factors separated in the analysis, to better assess their importance and contribute. By means of a visual analysis of the fourth graph, we were able to identify clear clusters of subjects that were involved in court trials: we propose using this network as an alarm trigger, suggesting to carefully check all the nodes in a cluster as soon as one of them gets involved in suspicious or illicit operations. In addition, the analysis of the first three networks gave evidence to the importance of studying structural holes and network centrality as ways to identify clients at a higher risk of fraud. Specifically, clients with higher risk profiles are less central in the transactions network (closeness centrality) and usually deal with financial operations of bigger amounts. This partially contrast with other studies (Ferrara et al., 2014) that pointed to higher centrality scores to identify criminals. However, since honest nodes are predominant in the network, we might argue that these nodes are more closely connected to one another, whereas criminals lie on more peripheral positions. This could also be an effect of new companies sometimes created to fulfill criminal intents, that relatively quickly can enter and leave the network, without having the time of acquiring more central positions. On the other hand, a larger in-degree centrality on the Geographical Area Network and a lower network constraint on the business sector graph prove to increase the probability of dealing with a subject involved in court trials. Therefore, we suggest to focus the attention on those clients who operate in more risky sectors and who are less constrained, i.e. who spread their connections across several regions and countries that pose a bigger threat to the stability of the financial system.

Our contribute has some similarities with the work of Drezewski et al. (2015), who use a SNA approach for money laundering detection, even if their primary focus is on criminal social roles, whereas ours is on network mapping and the identification of risk profiles of the companies involved in the factoring business. An advantage of our study is that we worked on real world data to generate models which are relatively easy to implement. However, the study present some limitations and could be further extended with future research. Firstly, a bigger sample could be analyzed, also involving different financial institutions, to check if our findings can be generalized or if they are more effective when related to factoring businesses. Secondly, it would be worth studying the effect of additional control variables that were not available for this research, such as the age and size of the companies involved in the financial operations. It might happen that some smaller and new firms are specifically created with the purpose of money laundering. Accordingly, one could also focus on the analysis of behaviors of leavers and newcomers in the networks. Our research could be enriched testing other dependent variables, for a more comprehensive definition of risk profiles. A main limitation of our dependent variable is the fact of being based on past events data. Additionally, one could consider other techniques to classify risky business sectors and geographical regions. Lastly, stochastic actor based models could be used to explain the mechanisms that drive the evolution of networks over time (Snijders, van de Bunt, & Steglich, 2010).

Our suggestion is that the metrics that come from our contribute should be combined with other tools – based, for instance, on machine learning algorithms, and possibly combining data at a national level – to allow a quick detection of suspicious nodes, with the final aim of preventing money laundering.

## Acknowledgements

We are grateful to Donato Franco, Senior Manager at KPMG Spa, for sharing his knowledge about anti-money laundering regulations in Italy and their practical applications. We thank Claudia Colladon for her help in revising this manuscript.

## References

Ahmed, M., Mahmood, A. N., & Islam, M. R. (2015). A survey of anomaly detection techniques in financial domain. *Future Generation Computer Systems*, *55*, 278–288.

Bose, I., & Mahapatra, R. K. (2001). Business data mining - A machine learning perspective. *Information and Management*, *39*(3), 211–225.

Burt, R. S. (1995). *Structural Holes: The Social Structure of Competition*. Cambridge, MA: Harvard University Press.

Burt, R. S. (2004). Structural Holes and Good Ideas. *American Journal of Sociology*, *110*(2), 349–399.

Carley, K. M. (2006). Destabilization of covert networks. *Computational and Mathematical Organization Theory*, *12*(1), 51–66.

Castellón González, P., & Velásquez, J. D. (2013). Characterization and detection of taxpayers with false invoices using data mining techniques. *Expert Systems with Applications*, *40*(5), 1427–1436.

De Nooy, W., Mrvar, A., & Batagelj, V. (2011). *Exploratory social network analysis with Pajek (2nd Ed.)*. Cambridge University Press.

Didimo, W., Liotta, G., Montecchiani, F., & Palladino, P. (2011). An advanced network visualization system for financial crime detection. In *2011 IEEE Pacific Visualization Symposium* (pp. 203–210). Hong Kong: IEEE.

DIRECTIVE 2005/60/EC OF THE EUROPEAN PARLIAMENT AND OF THE COUNCIL on the prevention of the use of the financial system for the purpose of money laundering and terrorist financing. (2005) OJ L 309/15.

Dreżewski, R., Sepielak, J., & Filipkowski, W. (2015). The application of social network analysis algorithms in a system supporting money laundering detection. *Information Sciences*, *295*, 18–32.

Everton, S. F. (2012). *Disrupting Dark Networks*. New York, NY: Cambridge University Press.

Ferrara, E., De Meo, P., Catanese, S., & Fiumara, G. (2014). Detecting criminal organizations in mobile phone networks. *Expert Systems with Applications*, *41*(13), 5733–5750.

Freeman, L. C. (1979). Centrality in social networks conceptual clarification. *Social Networks*, *1*, 215–239.

Gao, S., & Xu, D. (2009). Conceptual modeling and development of an intelligent agent-assisted decision support system for anti-money laundering. *Expert Systems with Applications*, *36*(2, Part 1), 1493–1504.

Gao, Z., & Ye, M. (2007). A framework for data mining-based anti-money laundering research. *Journal of Money Laundering Control*, *10*(2), 170–179.

Irwin, A. S. M., Choo, K. K. R., & Liu, L. (2012). Modelling of money laundering and terrorism financing typologies. *Journal of Money Laundering Control*, 15(3), 316–335.

Khan, N. S., Larik, A. S., Rajput, Q., & Haider, S. (2013). A Bayesian Approach for Suspicious

Financial Activity Reporting. *International Journal of Computers and Applications*, *35*(4), 181–187.

Khanuja, H. K., & Adane, D. S. (2014). Forensic Analysis for Monitoring Database Transactions. In J. L. Mauri, S. M. Thampi, D. B. Rawat, & D. Jin (Eds.), *Security in Computing and Communications* (pp. 201–210). Springer Berlin Heidelberg.

Kharote, M., & Kshirsagar, V. P. (2014). Data Mining Model for Money Laundering Detection in Financial Domain. *International Journal of Computer Applications*, *85*(16), 61–64.

Lopez-Rojas, E. A., & Axelsson, S. (2012). Money Laundering Detection using Synthetic Data. In *The 27th annual workshop of the Swedish Artificial Intelligence Society (SAIS)* (pp. 33–40). Örebro, Sweden: Linköping University Electronic Press.

Naheem, M. A. (2015). Trade based money laundering: towards a working definition for the banking sector. *Journal of Money Laundering Control*, *18*(4), 513–524.

Ngai, E. W. T., Hu, Y., Wong, Y. H., Chen, Y., & Sun, X. (2011). The application of data mining techniques in financial fraud detection: A classification framework and an academic review of literature. *Decision Support Systems*. Panigrahi, S., Kundu, A., Sural, S., & Majumdar, A. K. (2009). Credit card fraud detection: A fusion approach using Dempster–Shafer theory and Bayesian learning. *Information Fusion*, *10*(4), 354–363.

Pérez, D. G., & Lavalle, M. M. (2011). Outlier detection applying an innovative user transaction modeling with automatic explanation. In *2011 IEEE Electronics, Robotics and Automotive Mechanics Conference (CERMA)* (pp. 41–46). IEEE.

Rajput, Q., Khan, N. S., Larik, A., & Haider, S. (2014). Ontology Based Expert-System for Suspicious Transactions Detection. *Computer and Information Science*, *7*(1), 103–114.

REGULATION (EC) No 1893/2006. Establishing the statistical classification of economic activities NACE Revision 2 and amending Council Regulation (EEC) No 3037/90 as well as certain EC Regulations on specific statistical domains. (2006) OJ L 393/1.

Reuter, P., & Truman, E. (2004). *Chasing Dirty Money: Progress on Anti-Money Laundering*. Peterson Institute.

Roberts, N., & Everton, S. F. (2011). Strategies for Combating Dark Networks. *Journal of Social Structure*, *12*, 1–32.

Rohit, K. D., & Patel, D. B. (2015). Review On Detection of Suspicious Transaction In Anti-Money Laundering Using Data Mining Framework. *International Journal for Innovative*

*Research in Science & Technology*, *1*(8), 129–133.

Schott, P. A. (2006). *Reference Guide to Anti-Money Laundering and Combating the Financing of Terrorism* (Second Ed.). Washington, DC: The World Bank.

Scott, J., & Carrington, P. J. (2011). *The SAGE handbook of social network analysis*. London, UK: SAGE.

Sharman, J. C. (2008). Power and discourse in policy diffusion: Anti-money laundering in developing states. *International Studies Quarterly*, *52*(3), 635–656.

Simser, J. (2008). Money laundering and asset cloaking techniques. *Journal of Money Laundering Control*, *11*(1), 15–24.

Snijders, T. A. B., van de Bunt, G. G., & Steglich, C. E. G. (2010). Introduction to stochastic actor-based models for network dynamics. *Social Networks*, *32*(1), 44–60.

Sparrow, M. K. (1991). The application of network analysis to criminal intelligence: An assessment of the prospects. *Social Networks*, *13*(3), 251–274.

Tang, J., & Yin, J. (2005). Developing an intelligent data discriminating system of anti-money laundering based on SVM. In *2005 International Conference on Machine Learning and Cybernetics* (Vol. 6, pp. 3453 – 3457). Guangzhou, China: IEEE.

The Bank of Italy. (2014). *Annual Report Financial Intelligence Unit*. Rome, Italy.

Turban, E., Sharda, R. E., & Delen, D. (2015). *Decision support and business intelligence systems* (10th Editi). Pearson.

Wang, X., & Dong, G. (2009). Research on Money Laundering Detection Based on Improved Minimum Spanning Tree Clustering and Its Application. In *2009 Second International Symposium on Knowledge Acquisition and Modeling* (Vol. 2, pp. 62–64). IEEE.

Wasserman, S., & Faust, K. (1994). *Social Network Analysis: Methods and Applications.* New York, NY: Cambridge University Press.